\documentstyle[prc,aps]{revtex}     
\begin{document}                             
\preprint{IUCF-WMU-UWM-PITT}

\title{\LARGE Measurement of Spin Correlation coefficients \\
       in $\rm \vec p \vec p \rightarrow d\pi^+$}

\author{ 
B.\ v.\ Przewoski, J.T.\ Balewski\footnote{Permanent address:
  Institute of Nuclear Physics, Cracow 31342, Poland}, J.\ Doskow, H.O.\ Meyer,  \\ 
R.E.\ Pollock, T.\ Rinckel, P.\
Th\"orngren--Engblom\footnote{Present address: Department of Radiation
  Sciences, S - 75121 Uppsala, Sweden}, A.\ Wellinghausen \\
{\small \it Dept.\ of.\ Physics and Cyclotron Facility, 
 Indiana University, Bloomington, IN 47405} \\[1.2ex]
W.\ Haeberli, B.\ Lorentz$^{\ddag}$, F.\ Rathmann\footnote{Present address:  
    Forschungs Zentrum J\"ulich GmbH, 52425 J\"ulich, Germany.}, 
B.\ Schwartz and T.\ Wise \\
{\small \it University of Wisconsin-Madison, Madison, WI 53706} \\[1.2ex]
W.W.\ Daehnick and Swapan K.\ Saha\footnote{Permanent address: 
 Bose Institute, Calcutta 700009, India}\\
{\small \it Dept.\ of Physics and Astronomy, Univ.\ of Pittsburgh,
  Pittsburgh, PA 15260} \\[1.2ex]
P.V.\ Pancella \\
{\small \it Western Michigan University, Kalamazoo, MI 49008} }

\date{\today}
\maketitle

\begin{abstract}
The spin correlation coefficent combinations $\rm A_{xx}+A_{yy}$ and 
$\rm A_{xx}-A_{yy}$, the spin correlation coefficients $\rm A_{xz}$
and $\rm A_{zz}$, and the analyzing power were measured for 
$\rm \vec p \vec p \rightarrow d\pi^+$ between center-of-mass angles
$25^{\circ} \leq \theta \leq 65^{\circ}$ at beam energies of 350.5, 375.0
and 400.0~MeV. The experiment was carried out 
with a polarized internal target and a stored, polarized beam. 
Non-vertical beam 
polarization needed for the measurement of $\rm A_{zz}$ 
was obtained by the use of solenoidal spin rotators.
Near threshold, only a few partial waves contribute, and pion s- and
p-waves dominate with a possible small admixture of d-waves. Certain
combinations of the observables reported here are a direct measure of
these d-waves. The d-wave contributions are found to be 
negligible even at 400.0~MeV.

\end{abstract}
\vskip .2 in
\centerline{PACS numbers: 24.70.+s, 25.10.+s, 29.20.Dh, 20.25.Pj, 29.27.Hj}
\vskip .1 in

\narrowtext
\twocolumn

\section{Introduction}
Over the past fifty years NN$\rightarrow$NN$\pi$ reactions have 
received considerable interest. Of those, $\rm pp \rightarrow d\pi^+$
is probably the one which has been most extensively studied. 
This is because it is experimentally much easier to identify a
two-particle final state.
Most older measurements of this reaction are concentrated at higher
energies where production via the $\Delta$ resonance dominates. 
With the advent of storage ring technology and internal
targets the energy regime closer to
threshold has become accesible. The first NN$\rightarrow$NN$\pi$
measurements close to threshold were restricted to cross section and
analyzing power measurements [Ref.1-8], since polarized internal
targets were not yet available. 
Measurements of spin correlation coefficients close to threshold
became feasible only recently with the
availability of windowless and pure polarized targets internal to
storage rings [Ref. 9-11]. 

At energies well above threshold a number of measurements
of spin correlation coefficients in $\rm pp \rightarrow d\pi^+$
exist. Of these, the ones closest to threshold
are measurements of $\rm A_{zz}$  at 401 and 425~MeV [Ref. 12] 
which have been performed using an external beam and a polarized target. 
    
 A parametrization in terms of partial wave amplitudes of the
$\rm pp \rightarrow d\pi^+$ data from threshold to 580~MeV was 
carried out by Bugg
fifteen years ago [Ref. 13]. A more recent, updated  partial-wave analysis
is maintained by the Virginia group [Ref. 14].
With only the cross section and the analyzing power as input,
the number of free parameters is usually lowered by theoretical input
such as a constraint on the phases of the amplitudes which is provided
by Watson's theorem [Ref.15]. In this respect, a measurement of spin
correlation parameters represents crucial new information because one
can relate these observables to certain combination of amplitudes
 without any model assumptions.

     Close to threshold, s and p wave amplitudes in the pion channel
 with a possible small admixture of d waves, are
sufficient to parametrize the data. In the following, we will demonstrate
that combinations of the spin correlation observables presented here are
directly sensitive to the strength of these d waves.

\section{Experimental Arrangement}

In this paper we report measurements of spin correlation coefficients 
in $\rm \vec p \vec p \rightarrow d\pi^+$ at 350.5, 375.0 and 400.0~MeV
at center-of-mass angles between 25$^{\circ}$ and 65$^{\circ}$. 
The experiment was carried out at the Indiana Cooler with the 
PINTEX\footnote{{\bf P}olarized {\bf IN}ternal
{\bf T}arget {\bf EX}periment} setup. PINTEX is located in 
the A-region of the Indiana Cooler. In
this location the dispersion almost vanishes and the horizontal and
vertical betatron functions are small, allowing
the use of a narrow target cell.   
The target setup consists of an atomic beam source[Ref.16] which injects
polarized hydrogen atoms into the storage cell.

Vertically polarized protons from the cyclotron were stack-injected 
into the ring at 197~MeV,
reaching an orbiting current of several 100~$\mu$A within a few minutes. 
The beam was then accelerated. After typically 10 minutes of data
taking, the remaining beam was 
discarded, and the cycle was repeated. 

The target and detector used for this experiment are the same as
described in [Ref. 9], and a detailed account of the apparatus can be
found in [Ref. 17]. The internal polarized target consisted of an 
open-ended 25~cm long storage cell of 12~mm diameter and 25~$\mu$m wall 
thickness. The cell is coated with teflon to avoid depolarization 
of atoms colliding with the wall. 
During data taking, the target polarization $\vec{Q}$ is changed every
2~s pointing in sequence, up or down ($\pm$ y), left or right ($\pm$ x), 
and along or opposite to the beam direction ($\pm$ z). The magnitude of 
the polarization was typically $\vert \vec{Q} \vert \sim $ 0.75 and
is the same within $\pm$ 0.005 for all orientations [Ref. 18,19]. 

The detector arrangement consists of a stack of scintillators and wire
chambers, covering a forward cone between polar angles of 5 and
30$^{\circ}$. From the time of flight and the relative energy 
deposited in the layers of the detector, the outgoing charged particles 
are identified as pions, protons or deuterons.
The detector system was optimized for an experiment to study the 
spin dependence in  $\rm \vec p \vec p \rightarrow pp\pi^0$ and
$\rm \vec p \vec p \rightarrow pn\pi^+$ near
threshold [Ref. 9,10]
The $\rm \vec p \vec p \rightarrow d\pi^+$ data presented here are a 
by-product of that experiment.

Date were taken with vertical as well as longitudinal beam polarization.
To achieve non-vertical beam polarization the proton spin was
precessed by two spin-rotating
solenoids located in non-adjacent sections of the six-sided Cooler. 
The vertical and longitudinal components of the beam polarization 
$\vec{P}$ at the target are about equal, with a small sideways 
component, while its magnitude was typically 
$\vert \vec{P} \vert \sim $ 0.6. 
Since the solenoid fields are fixed in strength, the 
exact polarization direction depends on beam energy after acceleration. 
In alternating measurement cycles, the sign of the beam 
polarization is reversed. More details on the preparation of non-vertical 
beam polarization in a storage ring can be found in Ref. 19.

\section{Data acquisition and processing}
For each beam polarization direction,
data are acquired for all 12 possible polarization combinations of beam 
($+,-$) and target ($\pm$ x, $\pm$ y, $\pm$ z). The event trigger
is such that two charged particles are detected in coincidence. 
Then, a minimum $\chi^2$ fit to the hits in each of the four wire chamber  
planes is performed in order to determine how well the event conforms with 
a physical two-prong event that originates in the target.
Events with $\chi^2 \leq$ 5 were included in the final data sample.
The information from the $\chi^2$ fit allows us to determine the polar
and azimuthal angles of both charged particles. In the case of
a two-particle final state the two particles are coplanar and the
expected difference between the two azimuthal angles is $\Delta \phi$=180
$^{\circ}$. Events between 150$^{\circ} \leq \Delta \phi \leq$
210$^{\circ}$ are accepted for the final analysis. The polar 
angles of the pion and the deuteron are correlated such that the
deuteron, because of its mass, exits at angles close to the beam 
whereas pion laboratory  
angles as large as 180$^{\circ}$ are kinematically allowed.
At energies below 350.5~MeV the limited acceptance of the
detector setup is caused by deuterons travelling through the central 
hole of the detector stack, whereas at higher energies the angular
coverage is more and more restricted because of $\pi^+$ missing the
acceptance of 30$^{\circ}$.  
Events within 1$^{\circ}$  of the predicted   
correlation of the $\pi^+$-d polar scattering angles were included in the
final analysis. Since there is a kinematically allowed maximum deuteron
angle which depends on the beam energy, only those two-prong events
were included in the final data sample for which 
the deuteron reaction angle was $\leq$ 7$^{\circ}$, 8$^{\circ}$ and
9$^{\circ}$ at 350.5, 375.0, 400.0~MeV respectively. 
The correlation between the polar angles of the pion and the
deuteron and the coplanarity condition uniquely determine the 
$pp \rightarrow d\pi^+$ reaction channel. Therefore, the inclusion of
particle identification
gates in the analysis did not change the spin correlation coefficients
by more than a fraction of an error bar. Consequently, no particle
identification gates were used in the final analysis.  

Since an open-ended storage cell is used, the target is distributed
along the beam (z-) axis. The resulting target density is roughly
triangular
extending from $\rm z=-12$~cm to $\rm z=+12$~cm with z=0 being the 
location where
the polarized atoms are injected. The angular
acceptance of the detector depends on the origin of the event along
the z-axis.
Specifically, the smallest detectable angle
increases towards the upstream end of the cell. It was only possible
to detect deuterons which originated predominatly from the upstream 
part of the target. 
Since the kinematically allowed maximum deuteron 
angle increases with energy, vertex coordinates
between z=$-$12~cm and 4, 6, 8~cm at 350.5, 375.0 and 400.0~MeV, 
respectively, were accepted into the data sample. This way, background
events originating from the downstream cell walls were suppressed.

The known spin correlation 
coefficients of proton-proton elastic scattering [Ref. 20] are used to monitor 
beam and target polarization, concurrently with the acquisition of 
$\rm pp \rightarrow d\pi^+$ events. To this end, coincidences between two 
protons exiting near $\Theta_{lab} = 45^{\circ}$ are detected by two 
pairs of scintillators placed behind the first wire chamber at
azimuthal angles $\pm 45^{\circ}$ and $\pm 135^{\circ}$. From this 
measurement, the products $\rm P_yQ_y$ and $\rm P_zQ_z$ of beam and target 
polarization are deduced.
Values for the products of beam and target
polarization can be found in Table 1 of Ref. 11. 

Background arises from reactions in the walls of the target cell 
and from outgassing of the teflon coating [Ref. 17]. 
The contribution arising from a $\sim 1 \%$ impurity in the 
target gas is negligible. Background, which is not rejected by the 
software cuts described above, manifests itself as a broad distribution
underlying the peak at $\Delta \phi$=180$^{\circ}$.  
A measurement with a nitrogen 
target matches the shape of the $\Delta \phi$ distribution seen with the 
H target, except for the peak at 180$^{\circ}$. This shape is used to subtract 
the background under the $\Delta \phi$ peak.
Within statistics, the background shows no spin 
dependence and is independent of the polar angle. The subtracted
background ranges from 2$\%$ to 5$\%$.

\section{Determination of A$_y$, A$_{xx}$, A$_{yy}$ and A$_{xz}$}
The analyzing power and spin correlation coefficients were determined
using the method of diagonal scaling. Previously, we used diagonal scaling 
to analyze a series of experiments to measure spin correlation
coefficients in pp elastic scattering and a detailed desciption of the
formalism can be found in Ref. 21. 

Since we only measure
the {\it product} of beam and target polarization $\rm P \cdot Q$, we cannot
normalize beam and target analyzing power independently. 
Because of the identity of the colliding particles, beam and target
analyzing power as a function of center-of-mass angle 
are related such that $\rm A_y^b(\theta) =
-A_y^t(180^{\circ}-\theta)$, in particular  $\rm A_y^b(90^{\circ}) =
-A_y^t(90^{\circ})$. Consequently, for comparison with theory we use the  
quantity 
$\rm \sqrt{-(A^b_y(\theta) \cdot
  A^t_y(180^{\circ}-\theta))}$ 
for which the absolute 
normalization depends on P$\cdot$Q and therefore is known.
The final data with their statistical errors are shown in Fig.~1. 

\section{Partial waves}
        When one restricts the angular momentum of the pion to
l$_{\pi} \leq $2, there are seven partial waves possible, each corresponding
to a given initial and final state with all angular momentum quantum
numbers given. Using the usual nomenclature (see, e.g., Ref. 13), these
amplitudes are

\begin{equation} \begin{array}{lll}
a_0\ :\ ^1S_0 & \rightarrow\ ^3S_1,\ l_{\pi}\ & =\ 1 \\[1.2ex]
a_1\ :\ ^3P_1 & \rightarrow\ ^3S_1,\ l_{\pi}\ & =\ 0 \\[1.2ex]
a_2\ :\ ^1D_2 & \rightarrow\ ^3S_1,\ l_{\pi}\ & =\ 1 \\[1.2ex] 
a_3\ :\ ^3P_1 & \rightarrow\ ^3S_1,\ l_{\pi}\ & =\ 2 \\[1.2ex] 
a_4\ :\ ^3P_2 & \rightarrow\ ^3S_1,\ l_{\pi}\ & =\ 2 \\[1.2ex] 
a_5\ :\ ^3F_2 & \rightarrow\ ^3S_1,\ l_{\pi}\ & =\ 2 \\[1.2ex] 
a_6\ :\ ^3F_3 & \rightarrow\ ^3S_1,\ l_{\pi}\ & =\ 2 \ .
\end{array}
\end{equation}

        Very close to threshold, only s-wave pions are produced, and a
single amplitude (a$_1$) is sufficient to describe the reaction. Slightly
above threshold, the p-wave a$_2$ becomes significant, while the other
p-wave (a$_0$) remains small and is often neglected entirely. Among the 
d-wave amplitudes, at, for instance, 400~MeV, a$_6$ is the largest, 
followed by a$_4$, a$_3$, and a$_5$ with a factor of ~20 between a$_5$
and a$_6$.
This is known indirectly from partial-wave analyses (Ref. 13,14) 
of angular distributions of cross section and analyzing power. 
        The spin correlation data presented here offer a more direct 
study of the d-wave strength in $\rm pp \rightarrow d\pi^+$. 
In order to demonstrate this,
we need to express the observables in terms of the seven partial-wave
amplitudes (eq.1). This has previously been done for the cross section
and the analyzing power (Ref. 22). For the spin correlation coefficients,
we find analogously

\begin{equation} \begin{array}{ll}
\sigma & =\ {1 \over {16 \pi}} [a_0^2+a_1^2+a_2^2+C_1+(a_2^2-2 \sqrt{2}Re(a_0a_2^*) \\[1.2ex]
 & +B_1+C_2)P_2^0(cos\theta)+ 
  C_3P_4^0(cos \theta)] \\[1.2ex]
\end{array}
\end{equation}

\begin{equation} \begin{array}{ll}
\sigma & (A_{xx}+A_{yy}) =\  {1 \over {16 \pi}}
[-2a_0^2-2a_2^2+C_4+(-2a_2^2 \\[1.2ex]
 & +4 \sqrt{2}Re(a_0a_2^*)+C_5)P_2^0(cos \theta)+C_6P_4^0(cos \theta)]  \\[1.2ex]
\end{array}
\end{equation}

\begin{equation} \begin{array}{ll}
\sigma & (A_{zz}) =\ {1 \over {16 \pi}}
[-a_0^2+a_1^2-a_2^2+C_1-C_4+(-a_2^2 \\[1.2ex]
 & +2 \sqrt{2}Re(a_0a_2^*)+B_1+C_2-C_5)P_2^0(cos \theta)  \\[1.2ex] 
 & +(C_3-C_6)P_4^0(cos \theta)]   \\[1.2ex] 
\end{array}
\end{equation}

\begin{equation} \begin{array}{ll}
\sigma & (A_{xx}-A_{yy}) =\ {1 \over {16 \pi}}[(B_2+C_1)P_2^2(cos
\theta) \\[1.2ex]
 & +C_8P_4^2(cos \theta)],
\end{array}
\end{equation}

with $\sigma$ being the unpolarized differential cross section. 
The observables in Eqs.~2-5 are functions of the center-of-mass
reaction angle, $\theta$, and the functions $P_l^m(cos \theta)$ 
are the usual Legendre polynomials.
In the above equations,
the terms B$_n$ are caused by interference between s- and d-waves and
are given by a sum of terms Re(a$_1$ a$_i^*$), (i=3,4,5,6), weighted by 
numerical factors which follow from angular momentum coupling.
The terms C$_n$ contain d-waves only, i.e., a sum of a$_i^2$ and
Re(a$_i$ a$_k^*$) 
(i,k=3,4,5,6), again weighted by the appropriate numerical factors.
A derivation of eqs.2-5 can be found in Ref. 23. 

From eq.~5 one sees that the combination $\rm A_{xx}-A_{yy}$ 
vanishes at all angles when there are no d-waves. A departure from 
this behaviour would be caused by an interference between s- and d-waves.
It is also easy to see from  (eqs.~2-4) that for all reaction
angles, the following holds

\begin{equation}
A_{zz}-A_{xx}-A_{yy}=1-\delta
\end{equation}

where $\delta$ contains only C$_n$ terms, i.e., only d-wave amplitudes.

Thus, both quantities, $\rm A_{xx}-A_{yy}$ and 
$\rm A_{zz}-A_{xx}-A_{yy}-1$ are a direct measure
of d-wave contributions. Clearly, $\rm A_{xx}-A_{yy}$ provides a more 
sensitive test
because here d-waves interfere with the dominant a$_1$ amplitude.

\section{Final data and comparison with theory}
We compare our data to the two newest, published phase shift
solutions, namely

\begin{description}
\item[SP96] the published partial-wave analysis of the VPI group
           [Ref. 24]; range of validity 0-550 MeV, where the quoted energy
           is the laboratory energy of the pion in $\rm \pi^+ d
           \rightarrow p p$. For this solution a combined analysis of
           $\rm pp \rightarrow pp$, $\rm \pi d \rightarrow \pi d$ and 
             $\rm \pi^+ d \rightarrow p p$ was performed. In 
           particular, the overall phase in $\rm \pi^+ d \rightarrow p
           p$ was determined for this solution.
\item[BU93] the published partial-wave analysis of the VPI group and
          D.V. Bugg [Ref. 25]; range of validity 9-256 MeV.
\end{description}

Numerical values for both phase shift analyses 
have been obtained from the SAID interactive program [Ref. 14].
As can be seen from Fig.~1, our data are in good agreement with
both phase shift solutions, although the agreement with SP96
(overall $\chi^2$=0.90) is better than with BU93 
(overall $\chi^2$=1.29).
As a function of energy we have 
$\chi^2$=0.90, 0.87, 0.99 (SP96) and $\chi^2$=1.64, 0.88, 1.50 (BU93)
at 350.5, 375.0 and 400.0~MeV respectively. This indicates that the energy 
dependence of SP96 near threshold is slightly closer to the data.
Also shown in Fig.~1 are the $\rm A_{zz}$ data of Ref. 12 which
agree with our data. 

The quantity $\rm A_{xx}-A_{yy}$ provides a sensitive and direct test
of d-wave contributions in $\rm \vec p \vec p \rightarrow \rm d \pi^+ $. 
Our data are consistent with negligible d-wave contributions even at
400~MeV (see Fig.~1).

\section{summary}
In summary, we have measured the analyzing power 
and spin correlation coefficients
in $\vec p \vec p \rightarrow d\pi^+$. The spin correlation coefficients 
allow a direct determination of d-wave contributions. 
In particular, $\rm A_{xx}-A_{yy}$ provides a sensitive test
because here d-waves interfere with the dominant $a_1$ amplitude.
Our data are consistent with negligible d-wave contributions even at 400~MeV.
In addition, our data are well described by a partial wave solution
(SP96) of the Virginia group. 
Even at 

\begin{acknowledgements}
We are grateful for the untiring
efforts of the accelerator operations group at IUCF, in particular
G. East and T. Sloan. This work was supported in part by the National
Science Foundation and the Department of Energy. This work has been 
supported by the US National Science Foundation under Grants
PHY-9602872, PHY-9901529, PHY-9722556, and by the US Department of
Energy under Grant DE-FG02-88ER40438.
\end{acknowledgements}

\onecolumn
\pagebreak
\mediumtext

\input{epsf}
\epsfverbosetrue
\tolerance=1000
\baselineskip=14pt plus 1pt minus 1pt

\begin{figure}
\epsfxsize=15cm
\hspace{1.0cm}
\epsfbox{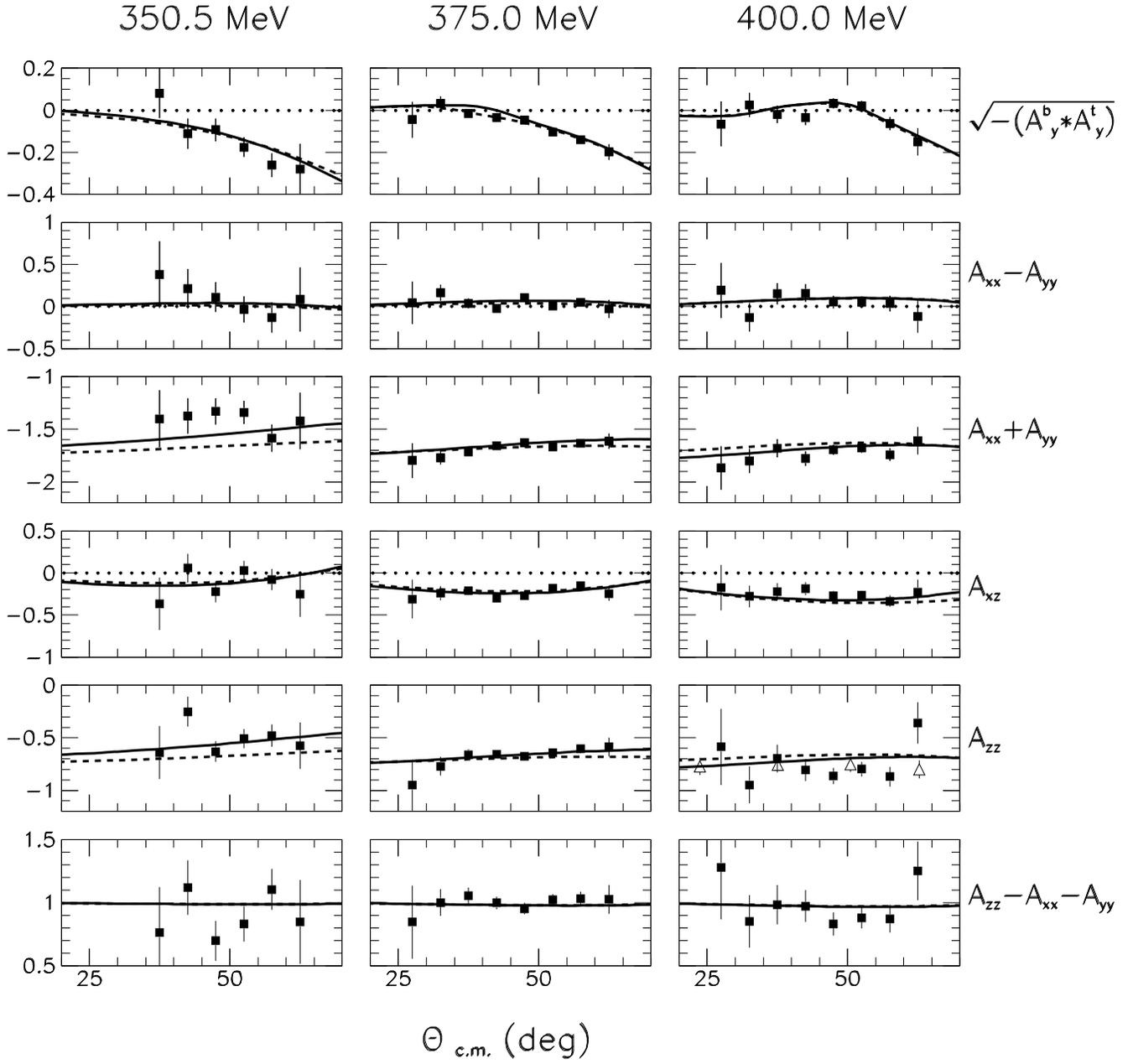}
\vspace{0.5cm}
\caption{Spin observables for $\rm \vec p \vec p \rightarrow d\pi^+$.
Beam and target analyzing power are taken at $\rm \theta$ and
$\rm 180^{\circ}-\theta$ respectively. 
The open triangles are  the $\rm A_{zz}$ measurement at 401 MeV of
Ref. 12. The solid and dashed lines are the SAID SP96 and SP93
solutions, respectively. The dotted line represents 0.}
\end{figure}


\begin{thebibliography}{16}
\bibitem{R01} P. Heimberg et al., Phys.\ Rev.\ Let.\ {\bf 77}, 96 (1996).

\bibitem{R02} Drochner et al.,  Phys.\ Rev.\ Let.\ {\bf 77}, 454 (1996).

\bibitem{R03} Korkmaz et al., Nucl.\ Phys.\ {\bf A535}, 637 (1991).

\bibitem{R04} Meyer et al., Nucl.\ Phys.\ {\bf A539}, 633 (1992).

\bibitem{R05} W.W. Daehnick et al. Phys.\ Rev.\ Let.\ {\bf 74}, 2913 (1995).

\bibitem{R06} J.G. Hardie et al. Phys.\ Rev.\ C\ {\bf 56}, 20 (1997).

\bibitem{R07} W.W. Daehnick et al. Phys.\ Let.\ B{\bf 423}, 213 (1998).

\bibitem{R08} R.W. Flammang et al. Phys.\ Rev.\ C\ {\bf 58}, 916 (1998).

\bibitem{R09} H.O. Meyer et al., Phys.\ Rev.\ Let.\ {\bf 81}, 3096 (1998).

\bibitem{R10} Swapan K. Saha et al., Phys.\ Let.\ B{\bf 461}, 175 (1999).

\bibitem{R11} H. O. Meyer et al. \ accepted for publication in Phys.\
  Rev.\ Let. 

\bibitem{R12} E. Aprile-Giboni et. al., Nucl.\ Phys.\ A{\bf 431}, 637 (1984). 

\bibitem{R13} D.V. Bugg, J. Phys. G: Nucl. Phys. 10, 47 (1984).

\bibitem{R14} The SAID dial-in: available via TELNET to
  clsaid.phys.vt.edu, with userid:said (no password required).

\bibitem{R15} K.M. Watson, Phys.\ Rev.\ {\bf 88}, 1163 (1952).

\bibitem{R16} T. Wise, A.D. Roberts and W. Haeberli, Nucl.\ Instr.\
  and Meth.\ {\bf A336} 410 (1993).

\bibitem{R17} T. Rinckel et al., Nucl.\ Instr.\ and\ Meth.\, accepted
  for publication. 

\bibitem{R18} F.\ Rathmann et al., \ Phys.\ Rev.\ C{\bf 58}, 658 (1998).

\bibitem{R19} B.\ Lorentz et al., \ Submitted to Phys.\ Rev.

\bibitem{R20} B. v. Przewoski et al., \ Phys.\ Rev.\ C{\bf 58}, 1897 (1998).

\bibitem{R21} H. O. Meyer, Phys.\ Rev.\ C{\bf 56}, 2074 (1997).

\bibitem{R22} C.L. Dolnick, Nucl.\ Phys.\ B{\bf 22}, 461 (1970).

\bibitem{R23} L.D. Knutson, Proc. 4$^{th}$ Int. Conf. on Nuclear
  Physics at Storage Rings, Bloomington, September 1999,
  eds. H.O. Meyer and P. Schwandt, AIP Conf. Proc., to be published.

\bibitem{R24} C.H. Oh et al., Phys.\ Rev.\ C{\bf 56}, 635 (1997).

\bibitem{R25} R.A. Arndt et al., Phys.\ Rev.\ C{\bf 48}, 1926 (1993).

  


\end{thebibliography}
\end{document}